 \documentstyle[12pt,graphicx]{article}  
\def\beq{\begin{equation}}
\def\eeq{\end{equation}}
\def\lsim{\ ^<\llap{$_\sim$}\ }
\def\gsim{\ ^>\llap{$_\sim$}\ }
\def\r2{\sqrt 2}
\def\beq{\begin{equation}}
\def\eeq{\end{equation}}
\def\beqn{\begin{eqnarray}}
\def\eeqn{\end{eqnarray}}

\def\PL{{1-\gamma_5\over 2}}
\def\PR{{1+\gamma_5\over 2}}
\def\sinW2{\sin^2\theta_W}

\def\mz2{M_{z}^2}
\def\c2b{\cos 2\beta}

\def\m#1{{\tilde m}_#1}

\def\mc#1{{\tilde m}_{\chi^{+}_#1}}

\def\mz{M_z}

\def\cb{\cos\beta}
\def\sb{\sin\beta}

\def\Fq2{F_{2}(q^2)}

\def\sec2w{sec^2\theta_W}

\def\gmin2{(g-2)_\mu}

\catcode`\@=11 

\def\lsim{\mathrel{\mathpalette\@versim<}}
\def\gsim{\mathrel{\mathpalette\@versim>}}
\def\@versim#1#2{\vcenter{\offinterlineskip
    \ialign{$\m@th#1\hfil##\hfil$\crcr#2\crcr\sim\crcr } }}

\def\PL{Phys. Lett.}

\def\PR{Phys. Rev.}
\def\ZPHY{{Z. Phys C} }

\begin{document}
\begin{flushright}
{CERN-TH/2001-119}\\
\end{flushright}
\begin{center}
{\Large\bf Slepton Flavor Nonuniversality, the Muon EDM
and its Proposed Sensitive Search at Brookhaven \\}
\vglue 0.5cm
{Tarek Ibrahim$^{(a)}$ and 
Pran Nath$^{(b,c)}$
\vglue 0.2cm
{\em 
$^a$ Department of Physics, Faculty of Science, University of 
Alexandria, Egypt}\\
{\em $^{(b)}$ Theoretical Physics Division, CERN CH 1211, Geneva, 
Switzerland\\}
{\em $^{(c)}$Department of Physics, Northeastern University, Boston,
MA 02115, USA\footnote{Permanent address}\\} }
\end{center}
\begin{abstract}
We analyze the electric dipole moment of the electron ($d_e$),
of the neutron ($d_n$) and of the muon ($d_{\mu}$) 
using the cancellation mechanism
  in the presence of nonuniversalities of the soft breaking parameters. 
It is shown that the nonuniversalities in the slepton sector
 produce a strong violation
of the scaling relation $d_{\mu}/d_e\simeq m_{\mu}/m_e$  in the
cancellation region. An analysis of  
$d_e, d_n$ and $d_{\mu}$ under the constraints
of the current experimental limits on $d_e$ and $d_n$ and under
the constraints of the recent Brookhaven result on $g_{\mu}-2$
  shows that in the non-scaling region $d_{\mu}$ can be as large 
  as ($10^{-24}-10^{-23}$)ecm and thus
  within reach of the recently proposed Brookhaven experiment for a 
 sensitive search for  $d_{\mu}$ at the level of $10^{-24}$ ecm.
 \end{abstract}

\section{Introduction}
Recently a proposal has been made to carry out a dedicated search
for the electric dipole moment of the muon at the level of
$10^{-24}ecm$\cite{sem1,sem2}. The current experimental limit 
on the muon edm 
from the previous CERN experiment is
$d_{\mu}<1.05\times 10^{-18}$ecm\cite{bailey}
and thus the proposed  experiment will improve the sensitivity for
$d_{\mu}$ by a factor of $10^5$ to $10^6$ over its previous measurement.
As is well known the electric dipole moment is an important probe
of new physics beyond the Standard Model. Thus the Standard Model
predicts very small values for the edms\cite{hoogeveen},
 e.g., $d_e<10^{-38}ecm$,
$d_{\mu}<10^{-35}ecm$ and $d_n<10^{-31}ecm$. However, many models 
of new physics predict much larger edms\cite{barr}
and thus an increased sensitivity of edm measurements is  
a probe of new physics. 
Supersymmetric (SUSY) models with softly broken supersymmetry are an 
example of this class of models as they provide abundant new
sources of CP violation since the soft breaking parameters can be
complex with CP phases O(1). 
In fact, with CP phases of O(1) an order of magnitude estimate shows
that the edm of  the electron and of the neutron 
are already in excess of the
experimental limits which are\cite{commins}

\beq
d_e<4.3\times 10^{-27}ecm,~~ d_n<6.3\times 10^{-26}ecm
\eeq
Several possible solutions have been offered to resolve this problem.
One possibility is to assume that the phases are small which, however,
 requires
a fine tuning\cite{ellis}. A second possibility is to assume that
the sparticles circulating in the loops are heavy, of masses in the
range of several TeV,
which, however, puts the sparticle spectrum beyond the reach 
of even the Large Hadron Collider (LHC)\cite{na}.   
Another alternative discussed more recently is the possibility of 
cancellations\cite{in2,inmssm}
 allowing for a satisfaction of the experimental limits
on the electron and on the neutron edms with
CP phases $O(1)$ and the sparticle spectrum within 
reach of accelerators. In addition to the above in certain specific
unified models, such as in left-right symmetric models, the troublesome
phases may automatically be small\cite{bdm1}.
 Now the normal expectation is that the
edms scale by the fermion mass and thus the muon edm should be
related to the electron edm by the scaling relation
$d_{\mu}\simeq(m_{\mu}/m_e)d_e$.  Using the experimental limit on
the electron edm of Eq.(1) this leads to $d_{\mu}<10^{-25}ecm$
which is below the sensitivity of the BNL experiment. 
However, we will show in this paper that the scaling relation 
$d_{\mu}\simeq (m_{\mu}/m_e)d_e$ can be violated strongly
 in the presence of nonuniversalities. The breakdown of scaling 
allows significant enhancement of the muon 
edm, by as much as a factor of 100, over the scaling result
and  $d_{\mu}\sim O(10^{-23})ecm$ can be gotten which is well
within reach of the proposed Brookhaven experiment.

The analysis presented  here is within the framework of MSSM.
Thus we do not assume any symmetries between the squark
and the slepton sector and consequently there are no constraints on the 
slepton sector arising from limits on flavor changing neutral
currents in the quark sector. Further, we do not impose the 
constraints from $\mu\rightarrow e\gamma$. This sector
of the theory involves additional parameters in MSSM and
thus no stringent constraints from this sector are expected
on the edm analysis. However, a similar analysis within a specific 
unified model will require
inclusion  of such constraints.
In addition to the mechanism discussed here, there exist 
previous analyses where a muon edm much larger than what is implied by
scaling can be generated. These consist of analyses in 
 left-right symmetric models\cite{ng},
 in a two Higgs doublet model\cite{barger}, and in models with certain
 assumed patterns of textures in the flavor space\cite{bdm2,babu}.
  The mechanism we
propose in this paper is different. Our technique is a natural extension of 
the cancellation mechanism using nonuniversalities which are already present
within MSSM. 

The outline of the rest of the paper is as follows: In Sec.2 we
review the basic formula for a charged lepton EDM in  MSSM.
 In the analysis we impose the constraints on
$a_{\mu}^{SUSY}$ from the BNL experiment\cite{brown} as a 
recent analysis has shown that the BNL result provides 
stringent constraints on
the phases\cite{icn}. In Sec.3 we give a discussion of the analysis 
and give results. Conclusions are given in Sec.4.

\section{Slepton nonuniversalities, $d_e$ and $d_{\mu}$}
The basic idea we use is that the cancellations arise
between the chargino and the neutralino exchange contributions
to the edm of the electron but such cancellations occur only
partially or negligibly for the  case of the edm of
the  muon. The  mechanism which brings this about arises from
nonuniversality between the electron and the muon sector. 
We give now the details of the analysis. First we 
review some basic formula that will be relevant in our discussion.
The chargino exchange contribution 
to the charged lepton edm is given by\cite{inmssm}
\beq
d_{\it l}(chargino)=\frac{e\alpha_{EM}}{4\pi \sin^2\theta_W}
\frac{\kappa_{\it l}}{m_{\tilde \nu_{\it l}}^2}
\sum_{1=1}^{2}\tilde m_{\chi_i^+}Im(U^*_{i2}V^*_{i1})
A(\frac{\tilde m_{\chi_i^+}^2}{m_{\tilde \nu_{\it l}}^2})
\eeq
where $\kappa_{\it l}$ is given by
$\kappa_{\it l}=m_{\it l}/(\sqrt 2 m_W \cos\beta)$,
 $A(x)=2(1-x)^{-2}(3-x+2lnx(1-x)^{-1})$,   and 
U and V in Eq.(2) are unitary transformations which diagonalize
the chargino mass matrix 

\beq
M_C=\left(\matrix{|\m2|e^{i\xi_2} & \r2 m_W  \sb \cr
	\r2 m_W \cb & |\mu| e^{i\theta_{\mu}}}
            \right)
\eeq
so that $U^* M_C V^{-1}=diag(\mc1,\mc2)$, 
 where $\theta_{\mu}$ is the phase of the Higgs mixing parameter
$\mu$ and $\xi_2$ is  the phase of the SU(2) gaugino mass $\tilde m_2$.
It is easily seen that the chargino exchange contribution to 
the edm of the charged lepton depends on just one combination of
the  phases, i.e., 
$\theta_{\mu}+\xi_2$.
The neutralino exchange contribution is given by\cite{inmssm} 
\beq
d_{\it l}(neutralino)=\frac{e\alpha_{EM}}{4\pi \sin^2\theta_W}
\sum_{k=1}^{2}\sum_{i=1}^{4} Im (\eta_{ik}^{\it l})
\frac{\tilde m_{\chi_i^0}}{M_{\tilde {\it l}_k}^2}
Q_{\tilde {\it l}}B(\frac{\tilde m_{\chi_i^0}^2}{M_{\tilde {\it l}_k}^2})
\eeq
where $B(x)=(2(x-1)^2)^{-1}(1+x+2xlnx(1-x)^{-1})$ 
and 
where $Q_{\tilde {\it l}}$ is the charge of the lepton and
  $\eta_{ik}^{\it l}$ is given by 
\beqn
\eta_{ik}^{\it l}=[-\sqrt 2\{\tan\theta_W(Q_{\it l}-T_{3\it l})X_{1i}
+T_{3\it l}X_{2i}\}D_{\it l1k}^*-\kappa_{\it l} X_{3i}D_{\it l2k}^*]\nonumber\\
(\sqrt 2 \tan\theta_W Q_{{\it l}} X_{1i}D_{{\it l} 2k}-\kappa_{{\it l}}
X_{3i}D_{{\it l}1k})
\eeqn
Here
 $X$ diagonalizes the  neutralino matrix $M_{\chi^0}$ so that
\beq
X^TM_{\chi^0}X=diag(\tilde m_{\chi_1^0}, \tilde m_{\chi_2^0},
\tilde m_{\chi_3^0},  
\tilde m_{\chi_4^0})
\eeq
where the matrix
$D_{\it l}$ diagonalizes the 
slepton (mass)$^2$ matrix so that 
$D_{\it l}^{\dagger} M_{\tilde{\it l}}^2D_{\it l}
=diag(M_{\tilde l_1}^2, M_{\tilde l_2}^2)$ where 
\beq
M_{\tilde {\it l}}^2=\left(\matrix{ M_{\tilde L_{\it l}}^2+m_{\mu}^2 -M_Z^2(\frac{1}{2}
-\sin^2\theta_W)\cos 2\beta & m_{\mu}(A_{\mu}^*m_0-\mu\tan\beta)\cr
m_{\mu}(A_{\mu}m_0-\mu^*\tan\beta) &   M_{\tilde R_{\it l}}^2+m_{\mu}^2 -M_Z^2\sin^2\theta_W
	\cos 2\beta  }
            \right)
\eeq
The neutralino exchange contribution depends additionally on
the phase combinations $\theta_{\mu}+\xi_1$, and 
$\theta_{\mu}+\alpha_{A_{\tilde {\it i}}}$. 
There are several ways in which the nonuniversality between the
selectron and the smuon channels can appear. For instance, the
sneutrino mass with the e flavor, i.e., $\tilde \nu_e$ may be 
different from the sneutrino mass for the $\mu$ flavor, i.e., 
$\tilde \nu_{\mu}$. In this case if one arranges cancellations 
between the chargino and the neutralino contributions to
occur for $d_e$  so that $d_e$ is in accord with the current experimental
limits, then such exact cancellations would not occur for the muon
channel due to different chargino contributions in this case because
of the disparity between the masses  $m_{\tilde \nu_e}$  and
$m_{\tilde \nu_{\mu}}$. A similar situation arises  because of the
nonuniversality in $M_{L\tilde e}^2$ and $M_{L\tilde \mu}^2$
and between $M_{R\tilde e}^2$ and $M_{R\tilde \mu}^2$.
In this case if the cancellation is arranged between the chargino and
the neutralino for $d_e$ it would be upset for $d_{\mu}$ 
due to neutralino contributions being different between the two cases.
There is yet a third possibility in which one may achieve a large
cancellation between the chargino and the neutralino contributions 
for $d_e$ but only a partial or no cancellation for $d_{\mu}$.
This consists in introducing nonuniversality only in  the
trilinear soft parameters in the selectron and the smuon 
channels, i.e., nonuniversalities between $A_e$ and $A_{\mu}$.
 In the following 
analysis we  focus on this type of nonuniversality. 
	In the analysis we impose the constraints
	from the recent results from Brookhaven\cite{brown}
	 on the size of new
	physics contributions to $a_{\mu}=(g_{\mu}-2)/2$.
	This experiment finds a deviation 
	from the Standard Model result\cite{czar}
	 at the $2.6\sigma$ level so that\cite{brown}

\beq
a_{\mu}^{exp}-a_{\mu}^{SM}=43(16)\times10^{-10}
\eeq
It has been known for some time that the supersymmetric electro-weak
contributions can make a significant correction to the muon g-2 and 
the size of this contribution can be as large or larger than the
Standard Model electro-weak correction\cite{kosower,lopez,chatto}.
It is also known that contributions to $g-2$ from models with
large extra dimensions are not significant and thus do not produce 
a strong background for supersymmetric contributions\cite{ny}
(see, however, Ref.\cite{desh}). Assuming then that all of the effect
of Eq.(8) comes from supersymmetry, one finds that the 
sparticle spectrum is severely constrained by the BNL result
and further that the sign of $\mu$ in the standard 
convention\cite{sugra} with CP invariance is determined to be 
positive\cite{chatto2} (Similar conclusions are drawn in 
Refs.\cite{kane,feng,gondolo}. See also Ref.\cite{marciano}).
A positive 
 $\mu$ sign is also the one favored by the $b\rightarrow s+\gamma$
constraint\cite{bsgamma}. 
Another important phenomenon is the effect of CP violating phases on
$a_{\mu}$.
It was shown in Ref.\cite{ing} that $a_{\mu}$ is a very
sensitive function of the CP phases\cite{ing} and that the CP phases
can change both the magnitude and the sign of  $a_{\mu}$.
 In the analysis
of Ref.\cite{icn} this sensitivity was used to  constrain 
the CP phases that enter $a_{\mu}$ by using the constraint of Eq.(8). 
It was found that as much as (60-90)\% of the parameter space of
CP phases may be eliminated by the BNL constraint\cite{icn}.
 The total contribution to
 $a_{\mu}$ arises from the chargino and neutralino exchanges
 where the  dominant contribution is from the chargino exchange 
 and in the presence of phases is given by\cite{ing}
\beqn
a^{\chi^{+
}}_{\mu}=\frac{m_{\mu}\alpha_{EM}}{4\pi\sin^2\theta_W}
\sum_{i=1}^{2}\frac{1}{M_{\chi_i^+}}Re(\kappa_{\mu} U^*_{i2}V^*_{i1})
F_3(\frac{M^2_{\tilde{\nu_{\mu}}}}{M^2_{\chi_i^+}})\nonumber\\
+\frac{m^2_{\mu}\alpha_{EM}}{24\pi\sin^2\theta_W}
\sum_{i=1}^{2}\frac{1}{M^2_{\chi_i^+}}
(|\kappa_{\mu} U^{*}_{i2}|^2+|V_{i1}|^2)
F_4(\frac{M^2_{\tilde{\nu_{\mu}}}}{M^2_{\chi_i^+}}).
\eeqn
Here
$F_3(x)$=${(x-1)^{-3}}$$(3x^2-4x+1-2x^2 lnx)$, 
  $F_4(x)$=${(x-1)^{-4}}$ $ (2x^3+3x^2-6x+1-6x^2 lnx)$ and
 $\kappa_{\mu}$ is defined as in Eq.(2).
As in the case of the edm analysis, it
is easily seen that the chargino exchange depends on the 
single phase  combination $\theta_{\mu}+\xi_2$, while 
 the neutralino contribution depends on several phase combinations,
 i.e.,  
 $\theta_{\mu}+\xi_i$ (i=1.2) and on $\theta_{\mu}+\alpha_{A_{\mu}}$.
 In carrying out the
analysis we use the $CP$ dependent formulae for $a_{\mu}$ 
and impose the constraints of Eqs.(1) and (8) on the phases.

\section{Analysis and Results} 
The general framework of our analysis is MSSM. However, since the
parameter space of this model is rather large we restrict our 
numerical analysis to a subset of parameters.
This subset consists of  $m_0$ (the universal scalar
mass at the GUT scale), $m_{\frac{1}{2}}$ (magnitude of the universal 
gaugino mass at the GUT scale), $|A_0|$ (the magnitude of the trilinear 
coupling of the soft parameters at the GUT scale for all sectors except
the smuon sector), and $\tan\beta =<H_2>/<H_1>$ where 
$H_2$ gives mass to the up quarks and $H_1$ gives mass to the
down quarks and the leptons. The magnitude of the Higgs mixing 
parameter $\mu$ 
is determined by the radiative breaking of the electro-weak symmetry.
We choose as independent the set of phases $\theta_{\mu}$, $\xi_i$ 
(i=1,2,3) (the phases of the U(1), SU(2) and SU(3) gauginos) and
$\alpha_{A_0}$, the phase of $A_0$. The trilinear soft parameters 
$A_t$, $A_b$, $A_e$ etc are all evolved from $A_0$ using renormalization 
group (RG) equations. Additionally in the smuon sector we
choose $|A_{\mu}|$ and $\alpha_{A_{\mu}}$ to be independent parameters
to generate the desired nonuniversalities in the smuon sector. 
The model described above is essentially a supergravity unified
model\cite{applied} with nonuniversalities\cite{nonuni} in the gaugino
 sector and in the slepton sector. 
Our technique consists in
first finding a set of soft  parameters with large phases which give
 a simultaneous cancellation for the case of the edm of the electron 
 and of the neutron. 
 Nonuniversalities are then used to generate violations of 
 scaling and a large edm for the muon.
 In the analysis we  include sbottom contribution to the
  gluonic dimension six operator in the case of the neutron and include
  two loop Higgs mediated
 contributions of the type discussed in Ref.\cite{twoloop}.
 Regarding the g-2 constraint we use a  2 sigma error corridor on
 Eq.(8) to constraint the sparticle spectrum and the phases, i.e., we use
 $10.6\times 10^{-10}<a_{\mu}^{SUSY}<76.2\times 10^{-10}$.

 In Table 1 we exhibit five cases
 each of which leads to a cancellation in the electron and in the neutron
 channel producing $d_e$ and $d_n$ in conformity with the current 
 data as shown in column 3 of Table 1. 
 For the muon case we then use all the same 
 parameters except $A_{\mu}$.  We consider three different cases here:
 (1) universal case where $A_{\mu}=A_e$, (2) $|A_{\mu}|=|A_e|$,
  $\alpha_{A_{\mu}}\neq \alpha_{A_e}$, and (3) $|A_{\mu}|\neq|A_e|$
  and $\alpha_{A_{\mu}}\neq \alpha_{A_e}$.
   Results of the analysis using these three cases are shown in Table 2, Table 3
   and Table 4 respectively. Thus for case(1) 
  one finds that the cancellations between the chargino and the
  neutralino contributions for $d_{\mu}$ parallel the cancellations
 of these contributions for $d_e$. Thus here a rough scaling results,
 i.e.,  $d_{\mu}\sim (m_{\mu}/m_e) d_e$ as can be seen by 
  comparing $d_e$ of Table 1 with $d_{\mu}$ of Table 2 
  (a detailed discussion of this phenomenon is given in the Appendix). 
  In this case the predicted values of $d_{\mu}$ 
  are smaller
  that than what the proposed Brookhaven experiment can probe.
  
   Next we discuss the other two cases which are more promising
   because they contain nonuniversality of the soft parameter
   A which leads to a breakdown of scaling and allows an
   enhancement of $d_{\mu}$ over what is allowed by scaling.
   (a detailed discussion of the breakdown of scaling 
   is given in the Appendix). 
  For case (2) the nonuniversality effects arise 
  from $\alpha_{A_{\mu}}\neq \alpha_{A_e}$ and the cancellation
  between the chargino and the neutralino contributions to
  $d_{\mu}$ do not parallel the cancellation  in $d_e$.
  Because of this one has only a partial cancellation in $d_{\mu}$
  in the region of the parameter space where one has a large 
  cancellation for $d_e$. This phenomenon leads to a violation
  of scaling and one obtains here values of $d_{\mu}$  much larger
  than expected from scaling.
  The results are exhibited in Table 3 and one finds that 
   for entries (a)-(e)  $d_{\mu}$  is typically larger
  than its corresponding value in Table 2 by a factor of 2-3.
  Further, all $d_{\mu}$ values listed in Table 3 
   fall in the range $O(10^{-24})$ecm which is within reach
  of  the proposed Brookhaven experiment\cite{sem1,sem2}.
Finally we  consider case(3) where 
   both the magnitude and the
  phase of $A_{\mu}$ differ from $A_e$. 
  In this case one can manage a much
  larger value for $d_{\mu}$ over the  value expected 
  from scaling. The results are shown in Table 4. In computing
 $d_{\mu}$ in this case we have used all the same parameters as
 in Table 1 except that the magnitude and the phase of $A_{\mu}$ are
 treated as independent. As can be seen from Table 4  one finds
 that $d_{\mu}$ can be as large as O($10^{-23}$)ecm in this case. 
 For the above three cases the $a_{\mu}$ listed in Tables (2), (3) and
 (4) is consistent with
 the  2 sigma error corridor constraint as discussed 
 at the end of the first paragraph in this section.

A plot of  $d_e, d_n$ as a function of $|A_0|$
 with the other parameters fixed by Table 1 is given
 in Figs. 1-5. Thus Fig.1 contains 
 case (a) of Table 1, Fig.2 contains case (b) of Table 1 etc.
The three cases, (1), (2) and (3) for $d_{\mu}$ given by Tables 2, 3
and 4 are also plotted. These figures thus illustrate graphically
the results
 discussed above. The analysis of Figs.1-5 shows that the 
 effect of  nonuniversality is to shift the point of maximum
 cancellation between the chargino and the neutralino contributions
  for $d_{\mu}$  relative to what happens in $d_e$. This shift 
 creates a much larger value for $d_{\mu}$ in the region
 where $d_e$ is compatible with the current experimental limits.
 \noindent
\begin{table}[h]
\begin{center}
\caption{{\bf EDMs of electron and of neutron
 for large phases}}
\begin{tabular}{|l|l|l|}
\hline
\hline
case &$\tan\beta$,$m_0$, $m_{\frac{1}{2}}$,$|A_e|$,
 $\xi_1$, $\xi_2$, $\xi_3$,$\theta_{\mu}$, $\alpha_{A_e}$ & 
 $d_e$, $d_n$ (ecm) \\
\hline
\hline
(a) &$10,150,281,0.78$,
$.5,-.45,.42,.2,4.1$ & $1.28\times 10^{-27}$, $-2.04\times 10^{-26}$  \\
\hline
\hline
(b) &$15,120,316,0.77$,
$-.6,-.15,-.64,.3,1.33$ & $3.15\times 10^{-27}$, $-4.8\times 10^{-27}$  \\
\hline
\hline
(c) &$20,200,246,0.93$,
$.28,-.51,-.11,.4,1.02$ & $-1.33\times 10^{-27}$, $-1.52\times 10^{-27}$  \\
\hline
\hline
(d) &$20,180,298,0.76$,
$-.6,-.07,-.49,.2,2.6$ & $3.45\times 10^{-27}$, $-2.24\times 10^{-27}$  \\
\hline
\hline
(e)&$6,100,246,1.31$,
$.4,-.77,.55,.4,1.15$ & $-3.8\times 10^{-27}$, $-3.9\times 10^{-26}$  \\
\hline
\hline
\end{tabular}
\end{center}
\end{table}

\noindent
\begin{table}[h]
\begin{center}
\caption{{\bf Muon edm for universal soft parameters}}
\begin{tabular}{|l|l|l|}
\hline
\hline
case&$|A_{\mu}|$,$\alpha_{A_{\mu}}$ & $d_{\mu}$ (ecm), $a_{\mu}$ \\
\hline
\hline
(a)&$.78,4.1$ & $2.5\times 10^{-25}$, $16.14\times 10^{-10}$\\
\hline
\hline
(b)&$.77,1.33$ & $6.47\times 10^{-25}$, $24.30\times 10^{-10}$\\
\hline
\hline
(c)&$.93,1.02$ & $-2.92\times 10^{-25}$, $34.85\times 10^{-10}$\\
\hline
\hline
(d)&$.76,2.6$ & $7.21\times 10^{-25}$, $28.50\times 10^{-10}$\\
\hline
\hline
(e)&$1.31,1.15$ & $-7.92\times 10^{-25}$, $13.04\times 10^{-10}$\\
\hline
\hline
\end{tabular}
\end{center}
\end{table}

\noindent
\begin{table}[h]
\begin{center}
\caption{{\bf Muon edm when $|A_{\mu}|=|A_e|$ and
$\alpha_{A_{\mu}}\neq \alpha_{A_e}$} }
\begin{tabular}{|l|l|l|}
\hline
\hline
case&$|A_{\mu}|$,$\alpha_{A_{\mu}}$ & $d_{\mu}$ (ecm), $a_{\mu}$ \\
\hline
\hline
(a)&$.78,-1.5$ & $1.07\times 10^{-24}$, $16.05\times 10^{-10}$\\
\hline
\hline
(b)&$.77,0.0$ & $1.17\times 10^{-24}$, $24.12\times 10^{-10}$\\
\hline
\hline
(c)&$.93,0.0$ & $1.51\times 10^{-24}$, $34.80\times 10^{-10}$\\
\hline
\hline
(d)&$.76,-2.0$ & $2.04\times 10^{-24}$, $28.30\times 10^{-10}$\\
\hline
\hline
(e)&$1.31,1.8$ & $-1.94\times 10^{-24}$, $13.27\times 10^{-10}$\\
\hline
\hline
\end{tabular}
\end{center}
\end{table}

\noindent
\begin{table}[h]
\begin{center}
\caption{{\bf Muon edm when $A_{\mu}\neq A_e$ }}
\begin{tabular}{|l|l|l|}
\hline
\hline
case&$|A_{\mu}|$,$\alpha_{A_{\mu}}$ & $d_{\mu}$ (ecm), $a_{\mu}$ \\
\hline
\hline
(a)&$7.0,1.5$ & $-1.33\times 10^{-23}$, $15.10\times 10^{-10}$\\
\hline
\hline
(b)&$8.0,1.0$ & $-1.27\times 10^{-23}$, $24.28\times 10^{-10}$\\
\hline
\hline
(c)&$6.0,-2.0$ & $-1.05\times 10^{-23}$, $35.80\times 10^{-10}$\\
\hline
\hline
(d)&$10.0,-2.5$ & $1.79\times 10^{-23}$, $29.02\times 10^{-10}$\\
\hline
\hline
(e)&$7.0,2.5$ & $-1.57\times 10^{-23}$, $14.40\times 10^{-10}$\\
\hline
\hline
\end{tabular}
\end{center}
\end{table}

\section{Conclusion}
In  this paper we have investigated the effect of nonuniversality 
in the trilinear soft parameter in the slepton sector on the
scaling relation $d_{\mu}/d_e\simeq m_{\mu}/m_e$.  We have shown
that large violations of this relation can occur 
 in the region of the parameter space where the cancellation
mechanism operates in the presence of nonuniversalities.  
Using this mechanism it is then shown that
values of $d_{\mu}$ as much as a factor of 
100 larger than implied by the scaling relation can be obtained.
Two branches of solutions were found. In the first solution 
the magnitudes $|A_{\mu}|$ and $|A_e|$ were assumed to be universal
and the nonuniversalities arose only in the phases of $A_e$ and
$A_{\mu}$, i.e.,$\alpha_{A_{\mu}}\neq\alpha_{A_e} $.
In this case violations of scaling were generally modest
but still sufficient to give $d_{\mu}$ of size $10^{-24}ecm$ 
consistent with Eqs.(1) and (8) and within reach of the 
sensitivity of the proposed BNL experiment\cite{sem1}.
In the second solution nonuniversalities both in the magnitude
and in the phase of the trilinear soft parameter were  assumed,
i.e., $|A_{\mu}|\neq|A_e|$ and  $\alpha_{A_e}\neq \alpha_{A_{\mu}}$.
In this case violations of the scaling formula were as much as a factor
of 100 and one can get $d_{\mu}\sim O(10^{-23})ecm$ consistent with
Eqs.(1) and (8). 
 Both of these possibilities are
 very encouraging for the observation of the muon edm in
 the proposed Brookhaven experiment  which can
 explore the muon edm at the level of $10^{-24}$ecm.
 Thus one may construe that the proposed Brookhaven experiment is an 
 important probe of both $CP$ violation and of nonuniversalities 
 in the muon sector.  \\
 
 \noindent
 \section{Acknowledgements}
A communication from Yannis  Semertzidis which led us to the present
investigation is acknowledged.
This research was supported in part by the NSF grant PHY-9901057.

\section{Appendix: The edm scaling relation 
 and its violation from nonuniversalities}
In this Appendix we explain the phenomenon seen in Tables 1-4 and in 
figs.1-5 where one finds that for the universal 
 values of $A$  the edm of the muon and of the electron satisfy 
a scaling relation, i.e., $d_{\mu}/d_e \simeq m_{\mu}/m_e$,
in the cancellation region,  while violations
of scaling occur due to nonuniversalities in A.
We exhibit this phenomenon for both chargino and neutralino
contributions.
 Now the chargino 
exchange contribution is strictly linear in the lepton mass as can be
seen from Eq.(2) and
we write $d_{\it l}(chargino)=m_{\it l} \bar d_{\it l}^C$ where
$\bar d_{\it l}^C$ is now independent of the charged lepton mass.
For the neutralino exchange contribution to $d_{\it l}$ there are
many types of terms that contribute.   The leading dependence of 
the lepton mass arises from $n_{ik}^l$ while subleading
dependence arises from the outside smuon mass factors in Eq.(4). 
Thus to understand the scaling phenomenon and its breakdown we
focus on $n_{ik}^l$ which can be expanded as follows using Eq.(5): 

\beqn
\eta_{ik}^{\it l}=a_0c_0X^2_{1i}D^*_{\it l 1k}D_{\it l 2k}+b_0c_0 X_{1i}X_{2i}D^*_{\it l 1k}D_{\it l 2k}
-\kappa_{{\it l}}a_0X_{1i}X_{3i}|D_{\it l 1k}|^2\nonumber\\
-\kappa_{{\it l}}b_0X_{2i}X_{3i}|D_{\it l 1k}|^2-
\kappa_{{\it l}}c_0X_{1i}X_{3i}|D_{\it l 2k}|^2
+\kappa^2_{\it l}X^2_{3i}D_{\it l 1k}D^*_{\it l 2k}
\eeqn
where $a_0$, $b_0$ and $c_0$ are independent of the lepton mass.
 The first  two terms on the right hand side of Eq.(10) 
 are linear in lepton
mass through the relation
\beqn
Im(D^*_{\it l11}D_{\it l21})=-Im(D^*_{\it l12}D_{\it l22})=\nonumber\\
\frac{m_{\it l}}{M^2_{\tilde{\it l1} }-M^2_{\tilde{\it l2}}}(m_0|A_{\it l}|\sin\alpha_f
+|\mu|\sin\theta_{\mu}\tan\beta).
\eeqn
The third, fourth and fifth terms on the right hand side of Eq.(10) have a leading linear 
dependence on the lepton mass
through the parameter $\kappa_{\it l}$
and have additional weaker dependence on the lepton mass through the
diagonalizing matrix elements $D_{ij}$. The last term in Eq.(10) 
is cubic in the 
lepton mass. However, in most of the parameter space cosidered, 
the first term in Eq.(10) is
the dominant one and controls the scaling behavior. Thus for the
case when all the soft SUSY
breaking parameters including A are universal (i.e., $A_l=A$) in Eq. (11)),
one finds from Eqs.(2), (4), (10) and (11),  
 that  scaling results, i.e.,   $d_{\mu}/d_e\simeq m_{\mu}/m_e$. 
This is borne out by the detailed 
numerical analysis as can be seen by comparing $d_{\mu}$ in Table 2
with $d_e$ in Table 1, and also by comparing $d_{\mu}$ for the
universal case (dotted line) with $d_e$ (solid line with
squares) in Figs.1-5.
 However, for the nonuniversal
case since the contribution from the A parameter is flavor dependent
we have a breakdown of
scaling here. This breakdown can be seen by comparing $d_{\mu}$ in Tables
3 and 4 with $d_e$ in Table 1, and also by comparing $d_{\mu}$ for the
nonuniversal cases (dashed line with triangles pointed down and
dashed line with triangles pointed up) with $d_e$ (solid line with
squares) in Figs.1-5.

\newpage
\begin{figure}[hbt]
\begin{center}
\includegraphics[angle=270,width=15cm]{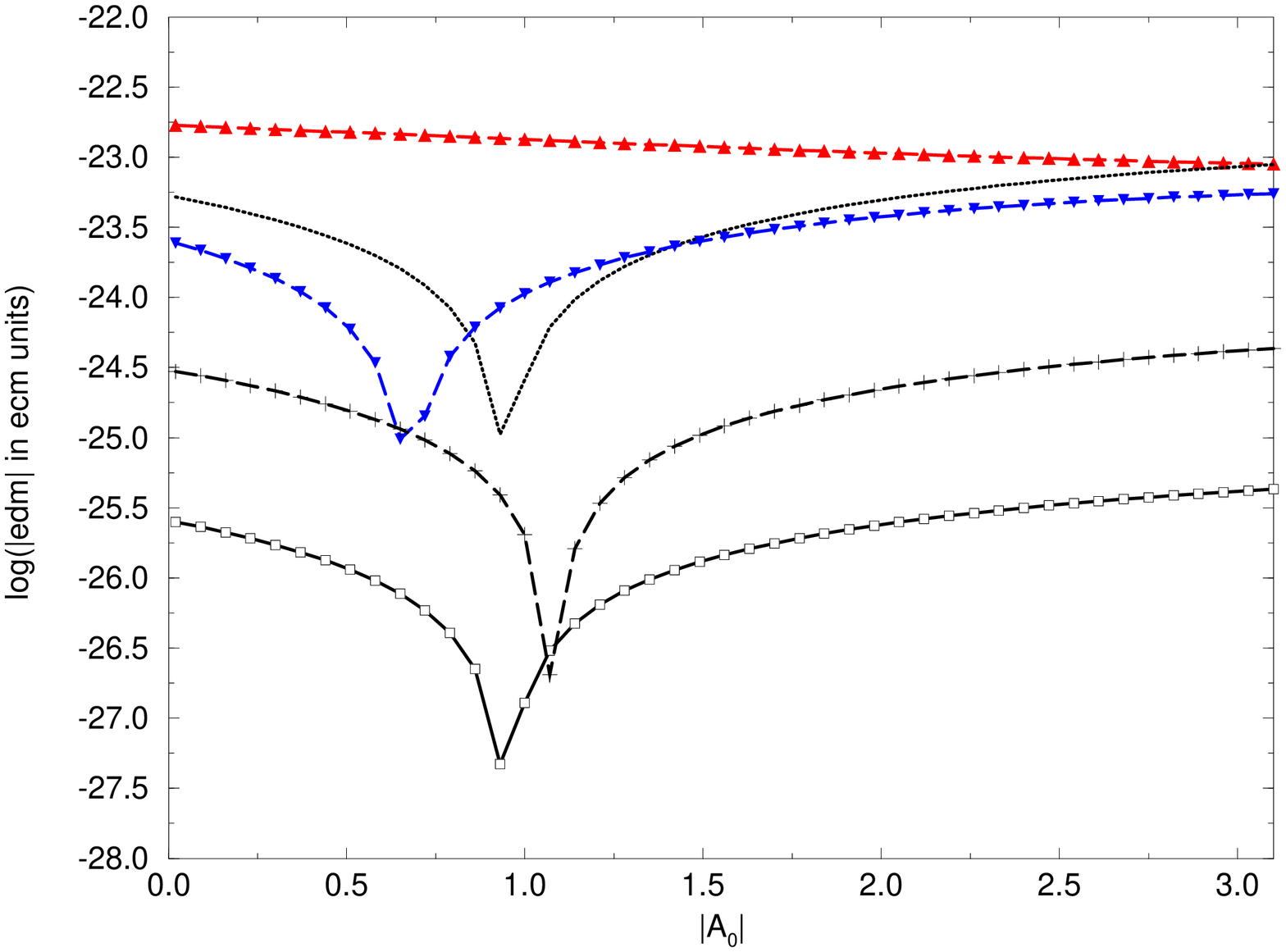}
\caption[]{A plot of the  electron edm $d_e$ (solid line with squares),
 the neutron edm $d_n$ (dashed line with plus signs), and the
 muon edm $d_{\mu}$ (dotted line) as a function of
 $|A_0|$ for the case when $\tan\beta =10$, 
$m_0=150$, $m_{\frac{1}{2}}=281$,  $\xi_1=.5$, 
$\xi_2=-.45$, $\xi_3=.42$, $\theta_{\mu}=.2$ and $\alpha_{A_e}=4.1$
where all masses are in GeV corresponding to case (a) in Table 1.
The curve with dashed line with triangles pointed down is a  plot of the
 muon edm $d_{\mu}$ which have all the same parameters as for 
 $d_e$ and $d_n$ except that $\alpha_{A_{\mu}}=-1.5$ (corresponding to
 case(a) of Table 3) and the 
 curve with dashed line with triangles pointed up is a  plot of the
 muon edm $d_{\mu}$ which have all the same parameters as for 
 $d_e$ and $d_n$ except that $|A_{\mu}|= 7.0$, and 
 $\alpha_{A_{\mu}}=-1.5$ (corresponding to
 case(a) of Table 4). }
\end{center}
\label{f1}
\end{figure}

\begin{figure}[hbt]
\begin{center}
\includegraphics[angle=270,width=15cm]{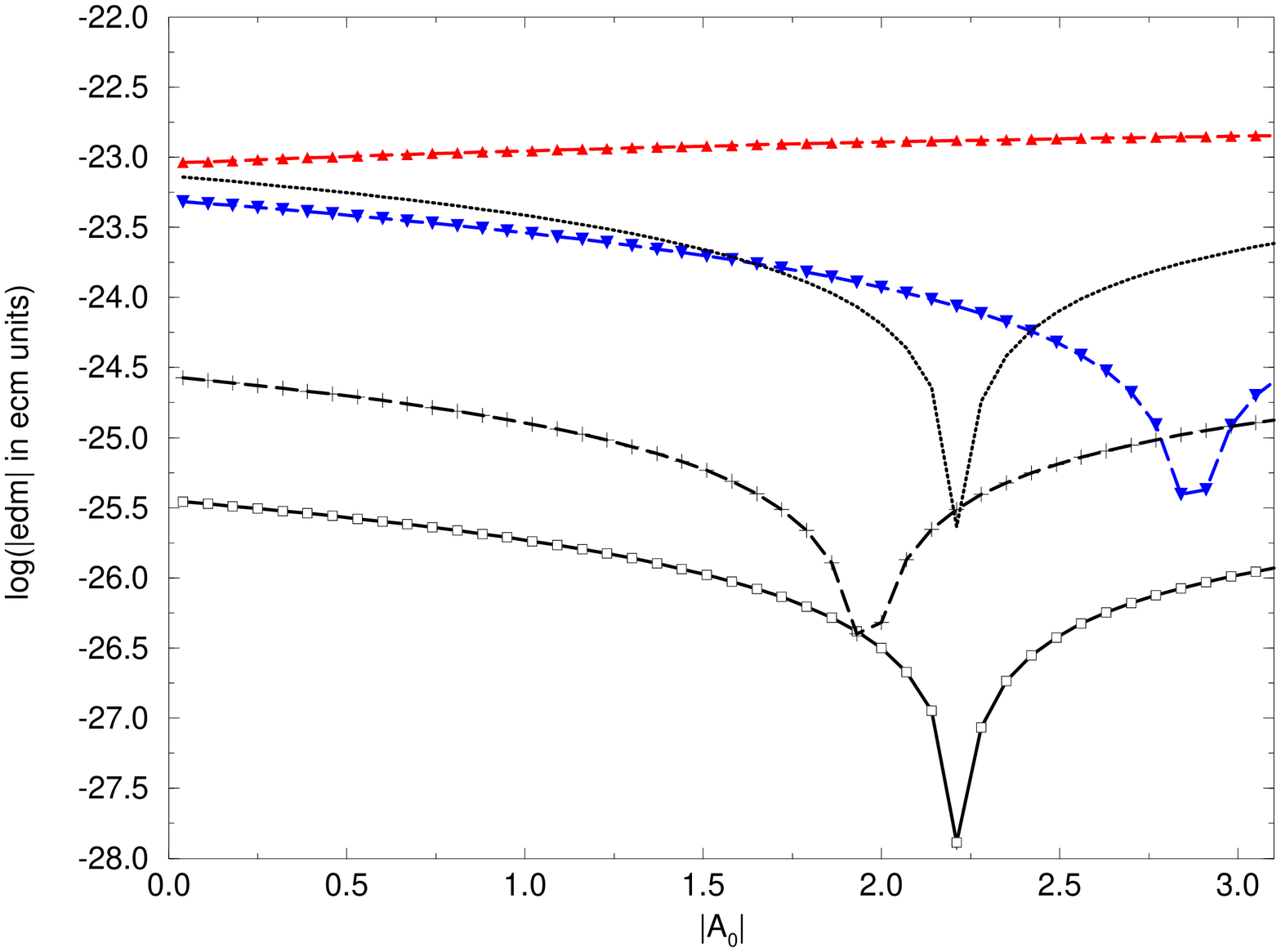}\\
\caption[]{A plot of the electron edm $d_e$ (solid line with squares),
of the neutron edm $d_n$ (dashed line with plus signs), and the muon
edm $d_{\mu}$ (dotted line)  
as a function of
 $|A_0|$ for the case when $\tan\beta =15$, 
$m_0=120$, $m_{\frac{1}{2}}=316$, $\xi_1=-.6$, 
$\xi_2=-.15$, $\xi_3=-.64$, $\theta_{\mu}=.3$ and $\alpha_{A_e}=1.33$
where all masses are in GeV corresponding to case (b) in Table 1.
The curve with dashed line with triangles pointed down is a  plot of the
 muon edm $d_{\mu}$ which have all the same parameters as for 
 $d_e$ and $d_n$ except that $\alpha_{A_{\mu}}=0.0$ (corresponding to
 case(b) of Table 3)and the 
 curve with dashed line with triangles pointed up is a  plot of the
 muon edm $d_{\mu}$ which have all the same parameters as for 
 $d_e$ and $d_n$ except that $|A_{\mu}|= 8.0$, and 
 $\alpha_{A_{\mu}}=1.0$ (corresponding to
 case(b) of  Table 4). }
\end{center}
\label{f2}
\end{figure}

\newpage
\begin{figure}[hbt]
\begin{center}
\includegraphics[angle=270,width=16cm]{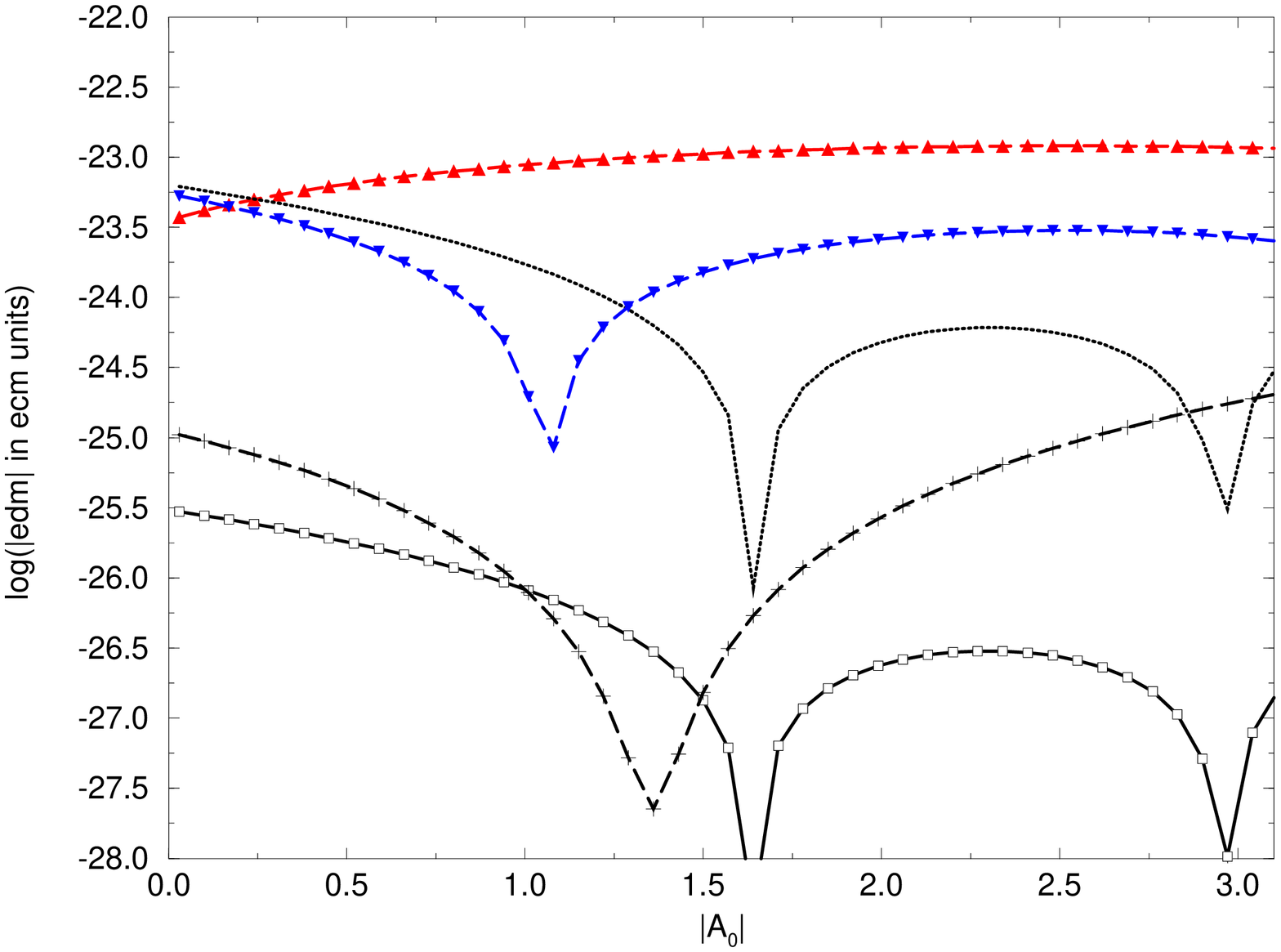}
\caption[]{A plot of the electron edm $d_e$ (solid line with squares), 
 of the neutron edm $d_n$ (dashed line with plus signs), and of the
 muon edm $d_{\mu}$  
as a function of
 $|A_0|$ for the case when $\tan\beta =20$, 
$m_0=200$, $m_{\frac{1}{2}}=246$, $\xi_1=.28$, 
$\xi_2=-.51$, $\xi_3=-.11$, $\theta_{\mu}=.4$ and $\alpha_{A_e}=1.02$
where all masses are in GeV corresponding to case (c) in Table 1.
The curve with dashed line with triangles pointed down is a  plot of the
 muon edm $d_{\mu}$ which have all the same parameters as for 
 $d_e$ and $d_n$ except that $\alpha_{A_{\mu}}=0.0$ (corresponding to
 case(c) of Table 3)and the 
 curve with dashed line with triangles pointed up is a  plot of the
 muon edm $d_{\mu}$ which have all the same parameters as for 
 $d_e$ and $d_n$ except that $|A_{\mu}|= 6.0$, and 
 $\alpha_{A_{\mu}}=-2.0$ (corresponding to
 case(c) of Table 4). }

\end{center}
\label{f3}
\end{figure}

\begin{figure}[hbt]
\begin{center}
\includegraphics[angle=270,width=16cm]{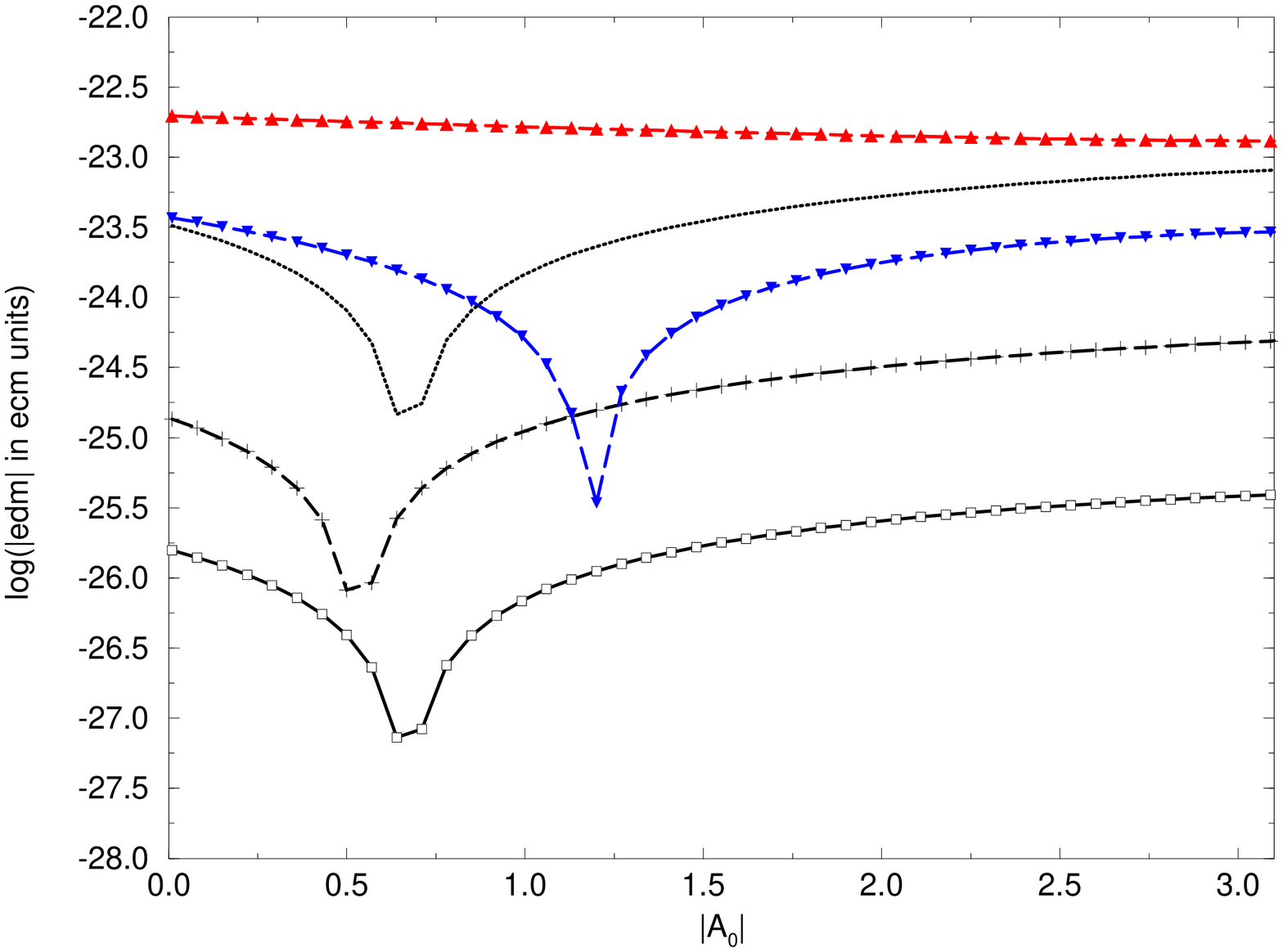}
\caption[]{A plot of the electron edm $d_e$ (solid line with squares),
 of the neutron edm $d_n$ (dashed line with plus signs),and the 
 muon edm $d_{\mu}$  
 as a function of
 $|A_0|$ for the case when $\tan\beta =20$, 
$m_0=180$, $m_{\frac{1}{2}}=298$, $\xi_1=-.6$, 
$\xi_2=-.07$, $\xi_3=-.49$, $\theta_{\mu}=.2$ and $\alpha_{A_e}=2.6$
where all masses are in GeV corresponding to case (d) in Table 1.
The curve with dashed line with triangles pointed down is a  plot of the
 muon edm $d_{\mu}$ which have all the same parameters as for 
 $d_e$ and $d_n$ except that $\alpha_{A_{\mu}}=-2.0$ (corresponding to
 (d) of Table 3)and the 
 curve with dashed line with triangles pointed up is a  plot of the
 muon edm $d_{\mu}$ which have all the same parameters as for 
 $d_e$ and $d_n$ except that $|A_{\mu}|= 10.0$, and 
 $\alpha_{A_{\mu}}=-2.5$ (corresponding to case (d) of 
 Table 4). }
\end{center}
\label{f4}
\end{figure}

\newpage
\begin{figure}[hbt]
\begin{center}
\includegraphics[angle=270,width=16cm]{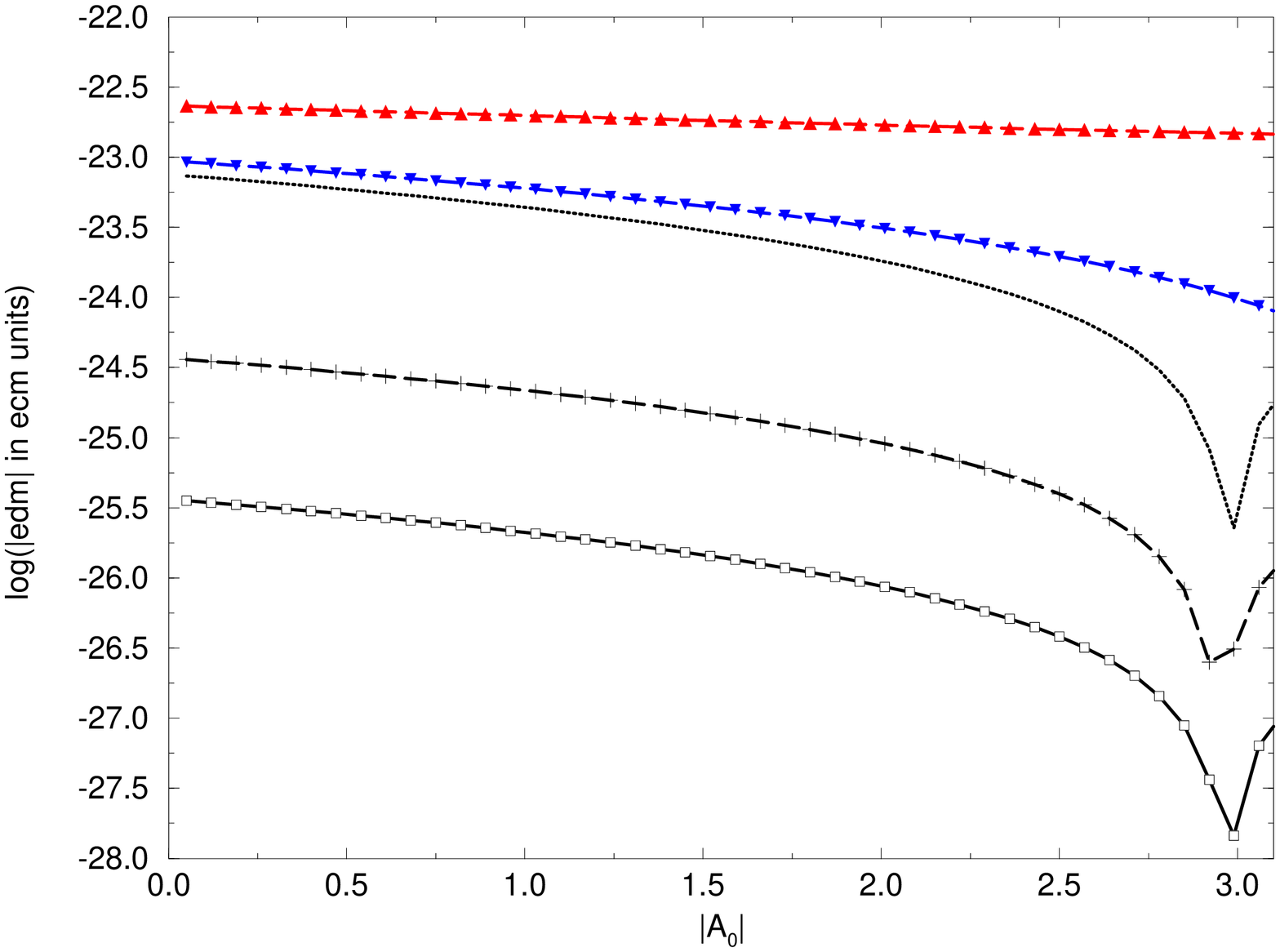}
\caption[]{A plot of the electron edm $d_e$ (solid line with squares),
 of the neutron edm $d_n$ (dashed line with plus signs),  and of the
 muon edm $d_{\mu}$ (dotted line) 
as a function of
 $|A_0|$ for the case when $\tan\beta =6$, 
$m_0=100$, $m_{\frac{1}{2}}=246$, $\xi_1=.4$, 
$\xi_2=-.77$, $\xi_3=.55$, $\theta_{\mu}=.4$ and $\alpha_{A_e}=1.15$
where all masses are in GeV corresponding to case (e) in Table 1.
The curve with dashed line with triangles pointed down is a  plot of the
 muon edm $d_{\mu}$ which have all the same parameters as for 
 $d_e$ and $d_n$ except that $\alpha_{A_{\mu}}=1.8$ (corresponding to
 (e) of Table 3)and the 
 curve with dashed line with triangles pointed up is a  plot of the
 muon edm $d_{\mu}$ which have all the same parameters as for 
 $d_e$ and $d_n$ except that $|A_{\mu}|= 7.0$, and 
 $\alpha_{A_{\mu}}=2.5$ (corresponding to case (e) of 
 Table 4). }
\end{center}
\label{f5}
\end{figure}

\end{document}